# Evaluation of Intel Max GPUs for CGYRO-based fusion simulations

Igor Sfiligoi

University of California San Diego, La Jolla, CA, USA

Jeff Candy and Emily A. Belli

General Atomics, La Jolla, CA, USA

Intel Max GPUs are a new option available to CGYRO fusion simulation users. This paper outlines the changes that were needed to successfully run CGYRO on Intel Max 1550 GPUs on TACC's Stampede3 HPC system and presents benchmark results obtained there. Benchmark results were also run on Stampede3 Intel Max CPUs, as well as NVIDIA A100 and AMD MI250X GPUs at other major HPC systems. The Intel Max GPUs are shown to perform comparably to the other tested GPUs for smaller simulations but are noticeably slower for larger ones. Moreover, Intel Max GPUs are significantly faster than the tested Intel Max CPUs on Stampede3.

## 1 INTRODUCTION

Fusion energy research has made significant progress over the years, yet the complexity of the turbulence in toroidal plasmas makes it difficult to accurately predict fusion reactor performance. While experimental methods are essential for gathering new operational modes, simulations are used to validate basic theory, plan experiments, interpret results on present devices, and ultimately to design future devices.

An important turbulence simulation tool is CGYRO [1], an Eulerian gyrokinetic solver designed and optimized for collisional, electromagnetic, multiscale simulations. CGYRO operates on a 6-dimensional grid (3D space + 2 D velocity + 1 D species), which allows for massive parallelism but also makes the simulation memory intensive. Most of the code is under CGYRO developers' control, with the only external dependency in time critical code being a FFT library.

The code was originally developed as a CPU-only solution but has been NVIDIA GPU-enabled since the deployment of the ORNL Titan HPC system [2] through the optional use of OpenACC directives. We had recently added support for OpenMP Target directives, too, as part of the effort to support AMD GPUs on the ORNL Frontier HPC system [3,4]. The CPU-only version relies on FFTW3 for computing FFTs, with cuFFT and hipFFT being used on the NVIDIA and AMD GPUs respectively.

The Stampede3 system [5,6] at the Texas Advanced Computing Center (TACC) is one of the first HPC systems to deploy Intel Max GPUs. This paper provides a description of the steps undertaken to enable CGYRO execution on those Intel GPUs, as well as benchmark results comparing it to both the Intel Max CPUs in a different partition of the same system and GPUs from other vendors on other HPC systems.

## 2 PORTING CGYRO TO INTEL MAX GPUS

On Stampede3, most of the CGYRO code compiled to Intel Max GPU binaries without any changes, by using the existing OpenMP Target directives and the Intel oneApi Fortran compiler [7]. The only code porting activity was related to the use of an external FFT library.

The obvious candidate library for performing FFT transforms on Intel Max GPUs was oneMKL [8]. The semantics of its batched multi-dimensional FFT interface is very similar to both cuFFT and hipFFT, which we were already using, so adding support was mostly trivial. The only major semantic difference is the reversal of the rank array order in the planning function. The function names between the various FFT libraries are instead different and rely on slightly different use of OpenMP directives, which we manage with the help of the compiler preprocessor. A subset of the relevant code showcasing those differences is available in Appendix A.

The default MPI library on Stampede3 is the Intel MPI library, so that's what we used. The Intel MPI library is GPU-aware, i.e. it can use GPU-resident memory buffers, but that functionality is disabled by default and must be explicitly enabled at runtime by setting `I_MPI_OFFLOAD=1`. The batch system is GPU-aware by default, properly pinning each MPI process to its own physical GPU, so no special setup was needed.

## 3 EVALUATING CGYRO ON INTEL MAX GPUS

### 3.1 HPC systems used for benchmarking

The Stampede3 system at TACC is comprised of several independent partitions. The `pvc` partition contains the Intel Max GPU nodes. Each node in that partition is equipped with four Intel Data Center GPU Max 1550 processors, also known as Ponte Vecchio Intel GPUs. Each node also has two Intel Xeon Platinum 8480 CPUs and 100Gbps Omni-Path networking.

The `spr` partition on Stampede3 contains HBM-based Intel Max CPUs, which were recently shown to be the most effective CPUs for CGYRO [9,10]. In particular, each node contains two Intel Xeon Max 9480 CPUs, paired with the same 100Gbps Omni-Path networking. While it was expected to be significantly slower on a per-node basis compared to the Intel Max GPUs, comparable benchmark results are included to allow for informed resource selection on that system.

We further benchmark CGYRO on two other modern GPU-based HPC systems, namely the Perlmutter `gpu&hbm80g` and `gpu` partitions at NERSC [11], and the Frontier HPC system [12] at the Oak Ridge National Laboratory (ORNL). The first is equipped with four NVIDIA A100 GPUs per node, while the latter is equipped with four AMD MI250X GPUS per node. The CPUs present in those nodes do not have significant influence on the benchmark results. Note, however, that compared to Stampede3 both systems have significantly more performant HPE Slingshot 11 networking delivering 800 Gbps per node, so while we provide the network communication costs of the benchmark simulations, we do not use them for comparison purposes.

From a software point of view, on the tested systems both the Intel Max GPU and the AMD MI250X GPU each appear as two logical accelerator devices. This setup avoids memory locality problems inherent to the chiplet nature of those GPUs. The Intel Max CPUs on Stampede3 do not expose the four chiplet setup to the user, but the batch system core binding achieves the same effect, thus avoiding cross-chiplet memory locality problems. The highly parallel nature of CGYRO simulation compute fits nicely in this partitioning scheme and we thus provide this information only for completeness. Table 1 provides a summary description of the 4 tested HPC partitions.

| HPC system | Main processor | Proc./node | Peak TFLOPs | Mem./proc (HBM2e) | Netw./node | FFT library |
|---|---|---|---|---|---|---|
| Stampede3 pvc | Intel Max1550 GPU | 4 (8 logical) | 52/proc (2 x 26) | 3.2 Tbps 128 GB (2 x 1.6 TBps 64 GB) | 1x 100 Gbps | oneMKL |
| Stampede3 spr | Intel Max 9480 CPU | 2 (8 NUMA) | 2.3/proc 4 x 0.6 | 1.6 Tbps 64 GB (4 x 0.4 TBps 16 GB) | 1 x 100 Gbps | MKL |
| Perlmutter gpu80 | NVIDIA A100 80 GB GPU | 4 (4 logical) | 9.7/proc | 2.0 TBps 80 GB | 4 x 200 Gbps | cuFFT |
| Perlmutter gpu | NVIDIA A100 40GB GPU | 4 (4 logical) | 9.7/proc | 1.6 TBps 40 GB | 4 x 200 Gbps | cuFFT |
| Frontier | AMD MI250X GPU | 4 (8 logical) | 48/proc 2 x 24 | 3.2 Tbps 128 GB (2 x 1.6 TBps 64 GB) | 4 x 200 Gbps | hipFFT |

Table 1. HPC systems used for CGYRO benchmarking.

### 3.2 CGYRO benchmark simulation inputs

CGYRO is a very versatile tools, allowing for simulations at multiple scales, from small linear ones to very large multiscale simulations. At the time of writing, Stampede3 gpu partition had a limit of 6 Intel Max GPU-based nodes per user job, so benchmarking had to be restricted to runs that could fit in less than 2 TB of memory, thus restricting full simulations to only small and medium sized parameters, named `nl02` and `sh03s` in the rest of the paper.

The largest amount of memory used by realistic simulation is reserved for holding the full collisional transform constants, which are problem specific, but can be computed once per simulation. CGYRO has the option of using a simplified collisional method that is less realistic but requires orders of magnitude less memory. Three additional benchmark simulation inputs with this simplified collisional mode are thus included to measure the impact of larger simulation grid sizes on the remaining code base.

Previous benchmarking activities have shown that FFT transforms are the dominant time consumer for CGYRO. The additional benchmarking input thus mimic the FFT sizes from largest tested full simulation up to those in recent leading-edge research [13]. These additional input parameters are called `nl03`, `bg03n`, `sh04n` and `bg04n` in the rest of the paper. Table 2 provides the relevant parameter values.

| Input | Collisional mode | Simulation grid size | 2D FFT size | Total FFTs batch |
|---|---|---|---|---|
| nl02 | Full (36 GB) | 38M = (192 x 24 x 32 x 16 x 8 x 2) | (288 x 96) | 6144 |
| sh03s | Full, fp32 (911 GB) | 425M = (480 x 32 x 48 x 24 x 8 x 3) | (720 x 144) | 18432 |
| nl03 | Simplified | 604M = (512 x 32 x 64 x 24 x 8 x 3) | (768 x 192) | 18432 |
| bg03n | Simplified | 573M = (864 x 24 x 96 x 18 x 8 x 2) | (1296 x 288) | 6912 |
| sh04n | Simplified | 906M = (1152 x 16 x 128 x 16 x 8 x 3) | (1728 x 384) | 6144 |
| ng04n | Simplified | 528M = (1344 x 16 x192 x 16 x 4 x 2) | (2016 x 576) | 2048 |

Table 2. Main relevant parameters of the CGYRO benchmark inputs

### 3.3 CGYRO benchmark results

All six CGYRO simulations have been benchmarked on both Intel GPUs and CPUs on Stampede3, on the AMD GPUs on Frontier and the 80GB NVIDIA GPUs on Perlmutter. Due to limited memory available on the 40GB NVIDIA GPUs on Perlmutter, only four of the six simulation inputs were benchmarked there. A total 28 data points were thus collected, and the complete timing information is available as tables in Appendix B, with raw logs also available in [14].

This section instead provides summary comparisons of the key benchmark outcomes. First, Fig. 1 provides relative performance numbers of the FFT dominated non-linear (nl) code section [2], which spends most of its time in the vendor-optimized FFT libraries. The Intel Max GPUs provide comparable performance to the other GPUs for small to medium sized FFT sizes but is significantly slower at larger FFT sizes. That said, the Intel Max GPUs are always an order of magnitude faster than the Intel Max CPUs.

The situation is slightly better on the CGYRO-maintained code base. As shown in Fig. 2, which provides relative aggregate performance numbers for those code sections, the Intel Max GPUs are only slightly slower than the AMD MI250X across all the tested parameter space. The situation is similar on NVIDIA GPUs for smaller problems, but the Intel Max GPUs are significantly slower than NVIDIA A100 80G GPUs on larger problems. Finally, the Intel Max GPUs are again an order of magnitude faster than the Intel Max CPUs on all tested inputs.

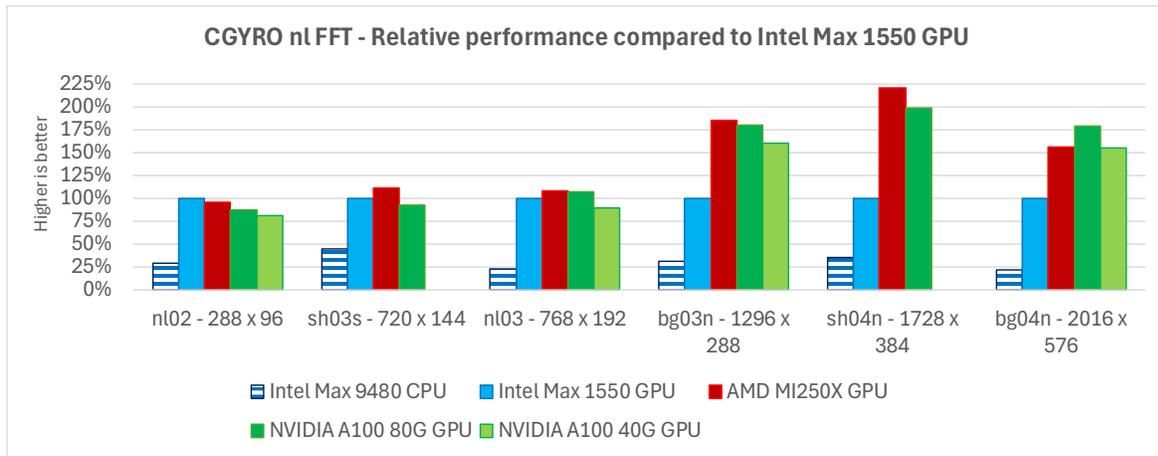

Fig 1. Relative performance on the FFT dominate nl code section

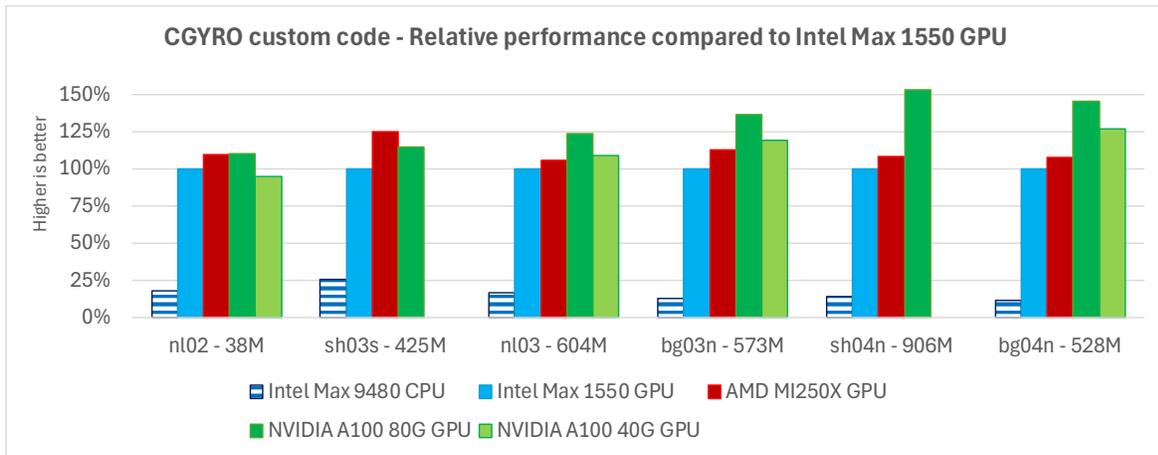

Fig. 2. Relative performance of the CGYRO-maintained code sections, excluding io and communication times

The markedly lower performance of the Intel Max GPUs is likely due to the inefficient use of its memory subsystem. Fig. 3 provides the benchmark information of the CGYRO code sections that are known to be completely memory bound. As can be seen, the Intel GPU is drastically slower compared to GPUs from other vendors on all inputs, and that discrepancy grows as the problem size grows. Our interpretation is that caching helps offset some of the performance deficiencies at smaller problem sizes, where a subset of the buffers fits completely into the GPU cache.

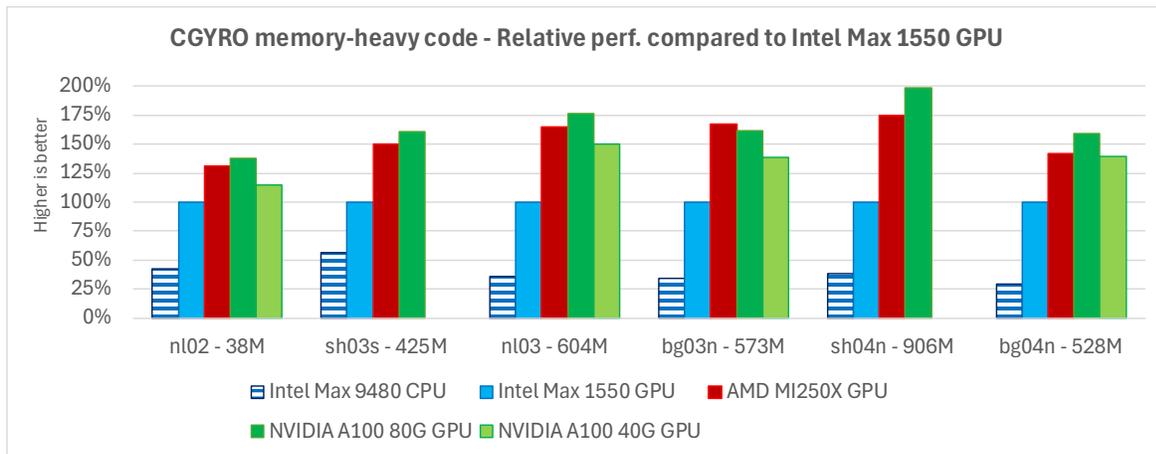

Fig. 3. Relative performance of the CGYRO memory-heavy code sections

Note that the previous figures do not include the time spent in inter-GPU communication, due to significantly different inter-node networking setup between the various systems. The comparison is however valid for single-node benchmark runs, so Fig. 4 provides the complete benchmark results for the `nl02` input. As can be seen, while still only a modest fraction of the total time, the time spent doing communication on Intel Max GPUs is about double compared to the other GPUs. This is compatible with the discrepancy in usable memory throughput observed in Fig. 3.

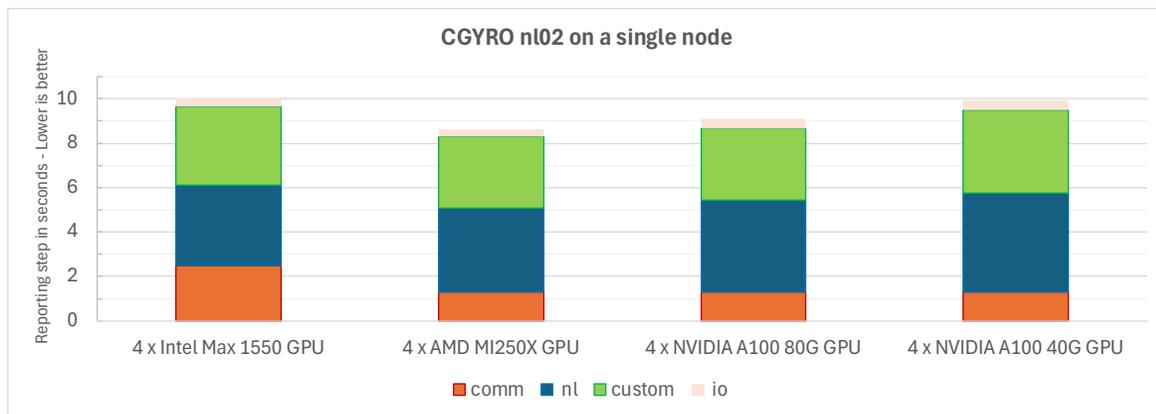

Fig. 4. Absolute time measured by reporting step for the CGYRO nl02 input, by benchmarked GPU system

## 4　SUMMARY AND CONCLUSIONS

The Intel Max GPUs have been shown to be a viable compute resource for CGYRO fusion simulations. Due to the use of OpenMP for handling compute offload, paired with FFT support in the oneMKL libraires, producing the binaries from the source code required only minor effort and is now fully supported in the mainstream CGYRO repository.

Performance wise, the Intel Max 1550 GPUs are almost on par with NVIDIA A100 and AMD MI250X GPUs on small to medium sized problems. On the larger problems, however, the Intel Max GPUs are noticeably slower, especially when computing FFT transforms, which are the dominant time consumer in CGYRO simulations. Unsurprisingly, the Intel Max 1550 GPUs are indeed drastically faster than the Intel Max 9480 CPUs.

In conclusion, the addition of Intel Max 1550 GPUs to the research computing ecosystem is a welcome addition. While their performance is not exceptional, it is still more than adequate. And the large amount of on-board memory further adds to their value.

**ACKNOWLEDGMENTS**

This work was partially supported by the U.S. Department of Energy under awards DE-FG02-95ER54309, DE-FC02-06ER54873, DE-SC0017992, and DE-SC0024425. Computing resources were provisioned from Stampede3 at the Texas Advanced Computing Center (TACC) through allocation PHY230202 from the Advanced Cyberinfrastructure Coordination Ecosystem: Services & Support (ACCESS) program, which is supported by National Science Foundation (NSF) grants #2138259, #2138286, #2138307, #2137603, and #2138296. An award of computing time was also provided by the ALCC and INCITE programs, and computing resources were provisioned at the National Energy Research Scientific Computing Center, which is an Office of Science User Facility supported under Contract DE-AC02-05CH11231, and the Oak Ridge Leadership Computing Facility, which is an Office of Science User Facility supported under Contract DE-AC05-00OR22725.

## APPENDIX A

Subset of the CGYRO 2D batched FFT planning code. Preprocessor directives are used to pick the right FFT library. Note that the Intel oneMKL library has a different semantics for `ndim` and `embed`, compared to cuFFT and hitFFT. The oneMKL library also requires the use of `omp dispatch`.

```fortran
      integer, dimension(2) :: ndim,inembed,onembed
#if defined(MKLGPU)
      ! oneMKL offload uses the reverse ordering
      ndim(2) = nx
      ndim(1) = ny
#else
      ndim(1) = nx
      ndim(2) = ny
#endif
      idist = ny2*nx
      odist = ny2*nx
      inembed = ny2
      onembed = ny2
#if defined(MKLGPU)
      inembed(2) = nx
      onembed(2) = nx
#endif

#if defined(MKLGPU)
      INTEGER*8 :: dfftw_plan = 0
!$omp target data map(tofrom: fymany,uymany)
!$omp dispatch
      call dfftw_plan_many_dft_c2r(dfftw_plan, 2, ndim, nffts,   &
             imany, inembed, 1, idist, &
             omany, onembed, 1, odist, FFTW_ESTIMATE)
!$omp end target data
#elif defined(HIPGPU)
      type(C_PTR) :: hip_plan = c_null_ptr
      istatus = hipfftPlanMany(hip_plan, 2, ndim, inembed, 1, idist, &
           onembed, 1, odist, HIPFFT_Z2D, nffts)
#else /* CUDAGPU */
      integer(c_int) :: cu_plan
      istatus = cufftPlanMany(cu_plan, 2, ndim, inembed, 1, idist, &
           onembed, 1, odist, CUFFT_Z2D, nffts)
#endif
```

Subset of the CGYRO batch FFT transform invocation. While the three supported libraries have similar semantics, they use different OpenMP mechanisms to attain GPU execution.

```
#if defined(MKLGPU)
!$omp target data map(tofrom: imany, omany)
#else
!$omp target data use_device_ptr(imany, omany)
#endif

#if defined(MKLGPU)
!$omp dispatch
  call dfftw_execute_dft_c2r(dfftw_plan, imany, omany)
#elif defined(HIPGPU)
  rc = hipfftExecZ2D(hip_plan, c_loc(imany), c_loc(omany))
#else /* CUDAGPU */
  rc = cufftExecZ2D(cu_plan, imany, omany)
#endif
!$omp end target data
```

# APPENDIX B

This section contains the raw measured times for the various CGYRO inputs. The timing sections are documented in [2].

Table B1. Measured time needed by reporting step for the CGYRO `nl02` input

| #XPUs (#Nodes) | XPU type | System | nl | coll | str | field | shear | mem | io | comm |
|---|---|---|---|---|---|---|---|---|---|---|
| 16 (8) | Intel Max 9480 CPU | Stampede3 | 3.1 | 1.8 | 2.3 | 0.5 | 0.0 | 0.3 | 0.1 | 5.2 |
| 4 (1) | Intel Max 1550 GPU | Stampede3 | 3.6 | 1.1 | 1.1 | 0.8 | 0.0 | 0.6 | 0.3 | 2.5 |
| 4 (1) | AMD MI250X GPU | Frontier | 3.8 | 0.8 | 1.5 | 0.5 | 0.0 | 0.4 | 0.3 | 1.3 |
| 4 (1) | NVIDIA A100 80G GPU | Perlmutter | 4.2 | 1.2 | 1.2 | 0.4 | 0.0 | 0.4 | 0.4 | 1.3 |
| 4 (1) | NVIDIA A100 40G GPU | Perlmutter | 4.5 | 1.5 | 1.4 | 0.4 | 0.0 | 0.5 | 0.4 | 1.3 |

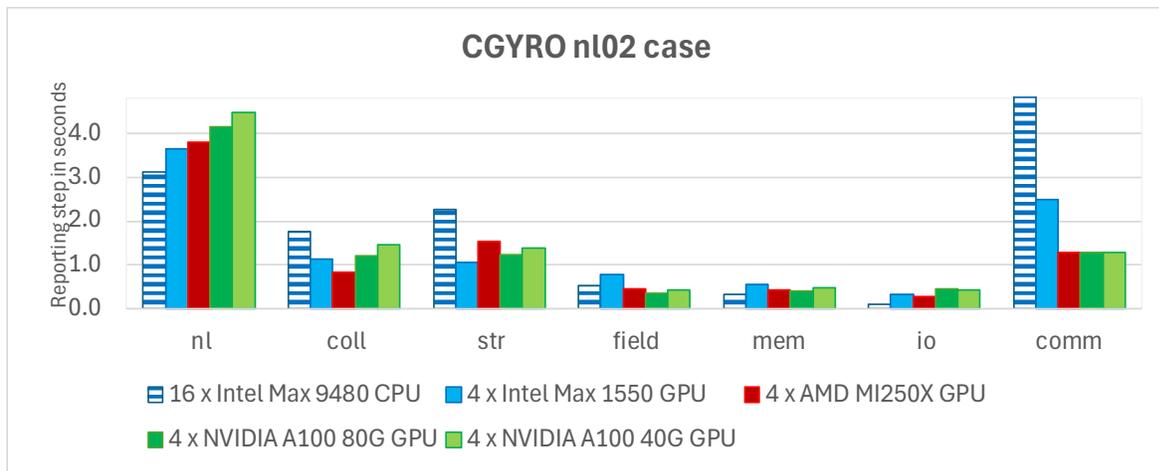

Fig B1. Comparison of measured times by code section for CGYRO `nl02` input

Table B2. Measured time needed by reporting step for the CGYRO `sh03s` input

| #XPUs (#Nodes) | XPU type | System | nl | coll | str | field | shear | mem | io | comm |
|---|---|---|---|---|---|---|---|---|---|---|
| 48 (24) | Intel Max 9480 CPU | Stampede3 | 16.1 | 6.8 | 9.8 | 2.2 | 3.6 | 1.7 | 0.2 | 35.5 |
| 24 (6) | Intel Max 1550 GPU | Stampede3 | 14.4 | 3.6 | 4.6 | 1.9 | 0.5 | 1.9 | 0.7 | 36.9 |
| 24 (6) | AMD MI250X GPU | Frontier | 12.9 | 2.3 | 5.4 | 0.8 | 0.3 | 1.3 | 0.4 | 10.4 |
| 24 (6) | NVIDIA A100 80G GPU | Perlmutter | 15.5 | 3.6 | 4.4 | 1.1 | 0.7 | 1.2 | 1.0 | 12.3 |

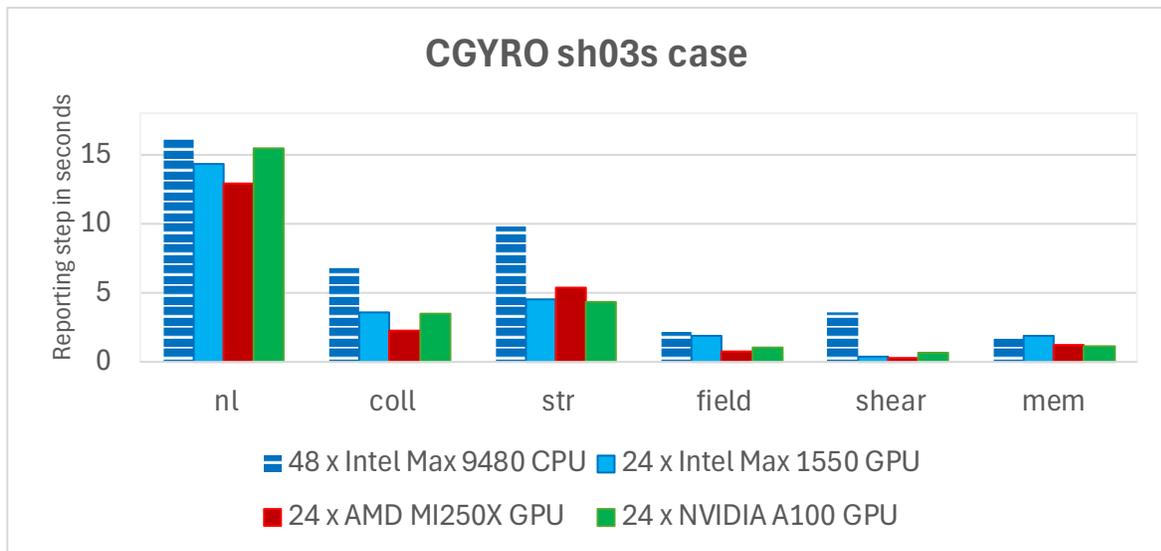

Fig B2. Comparison of measured times by code section for CGYRO `sh03s` input

Table B3. Measured time needed by reporting step for the CGYRO `nl03` input

| #XPUs (#Nodes) | XPU type | System | nl | coll | str | field | shear | mem | io | comm |
|---|---|---|---|---|---|---|---|---|---|---|
| 64 (32) | Intel Max 9480 CPU | Stampede3 | 17.1 | 1.4 | 9.6 | 2.2 | 0.0 | 1.7 | 0.2 | 22.8 |
| 16 (4) | Intel Max 1550 GPU | Stampede3 | 15.6 | 0.7 | 4.6 | 2.2 | 0.0 | 2.4 | 0.8 | 44.2 |
| 16 (4) | AMD MI250X GPU | Frontier | 14.4 | 1.3 | 5.9 | 0.8 | 0.0 | 1.5 | 0.8 | 12.0 |
| 16 (4) | NVIDIA A100 80G GPU | Perlmutter | 14.6 | 0.8 | 4.8 | 1.1 | 0.0 | 1.4 | 1.7 | 14.1 |
| 16 (4) | NVIDIA A100 40G GPU | Perlmutter | 17.4 | 0.8 | 5.5 | 1.3 | 0.0 | 1.6 | 1.8 | 15.0 |

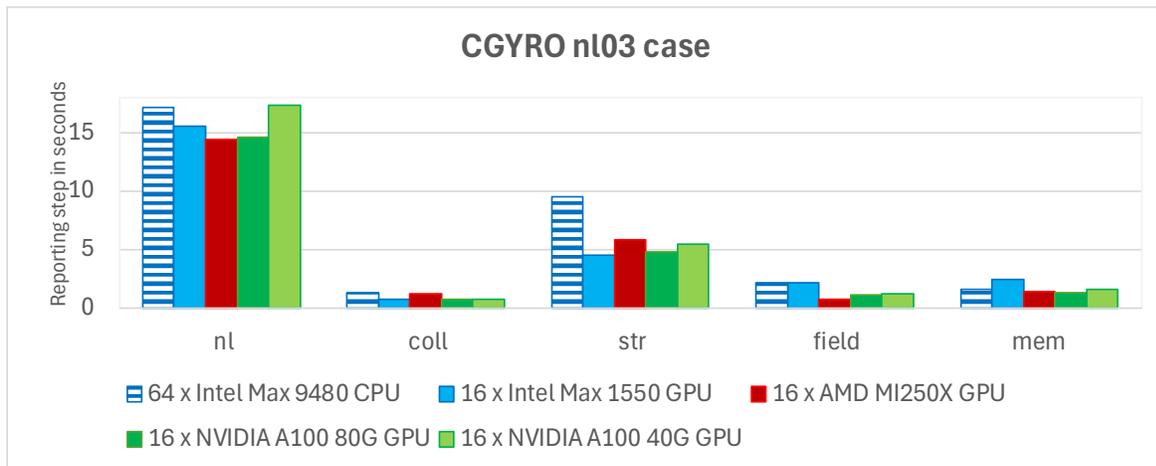

Fig B3. Comparison of measured times by code section for CGYRO `nl03` input

Table B4. Measured time needed by reporting step for the CGYRO `bg03n` input

| #XPUs (#Nodes) | XPU type | System | nl | coll | str | field | shear | mem | Io | comm |
|---|---|---|---|---|---|---|---|---|---|---|
| 64 (32) | Intel Max 9480 CPU | Stampede3 | 29.5 | 0.9 | 14.3 | 4.1 | 5.8 | 2.4 | 0.2 | 32.3 |
| 16 (4) | Intel Max 1550 GPU | Stampede3 | 36.4 | 0.5 | 7.2 | 2.8 | 0.7 | 3.3 | 0.7 | 49.3 |
| 16 (4) | AMD MI250X GPU | Frontier | 19.7 | 0.7 | 8.7 | 1.0 | 0.5 | 2.0 | 1.4 | 15.9 |
| 16 (4) | NVIDIA A100 80G GPU | Perlmutter | 20.2 | 0.4 | 6.0 | 1.1 | 1.2 | 2.1 | 1.2 | 22.7 |
| 16 (4) | NVIDIA A100 40G GPU | Perlmutter | 22.8 | 0.3 | 6.8 | 1.2 | 1.5 | 2.4 | 1.1 | 23.8 |

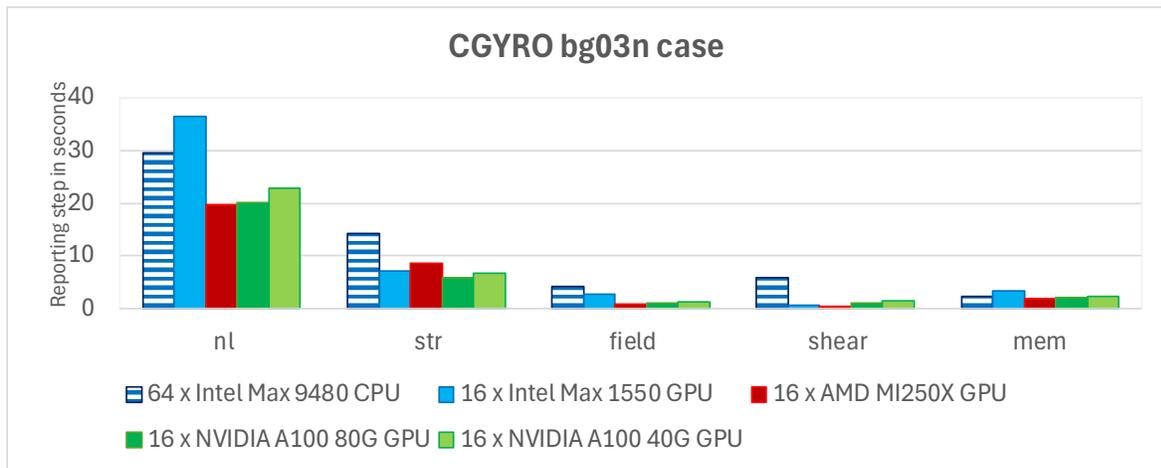

Fig B4. Comparison of measured times by code section for CGYRO `bg03n` input

Table B5. Measured time needed by reporting step for the CGYRO `sh04n` input

| #XPUs (#Nodes) | XPU type | System | nl | coll | str | field | shear | mem | io | comm |
|---|---|---|---|---|---|---|---|---|---|---|
| 64 (32) | Intel Max 9480 CPU | Stampede3 | 34.2 | 1.6 | 13.8 | 3.9 | 6.1 | 2.6 | 0.3 | 36.4 |
| 16 (4) | Intel Max 1550 GPU | Stampede3 | 47.6 | 0.8 | 7.0 | 3.1 | 0.7 | 4.0 | 1.0 | 70.6 |
| 16 (4) | AMD MI250X GPU | Frontier | 21.6 | 1.5 | 8.9 | 1.3 | 0.5 | 2.3 | 1.6 | 18.8 |
| 16 (4) | NVIDIA A100 80G GPU | Perlmutter | 24.0 | 0.5 | 5.6 | 1.4 | 0.8 | 2.0 | 1.3 | 30.4 |

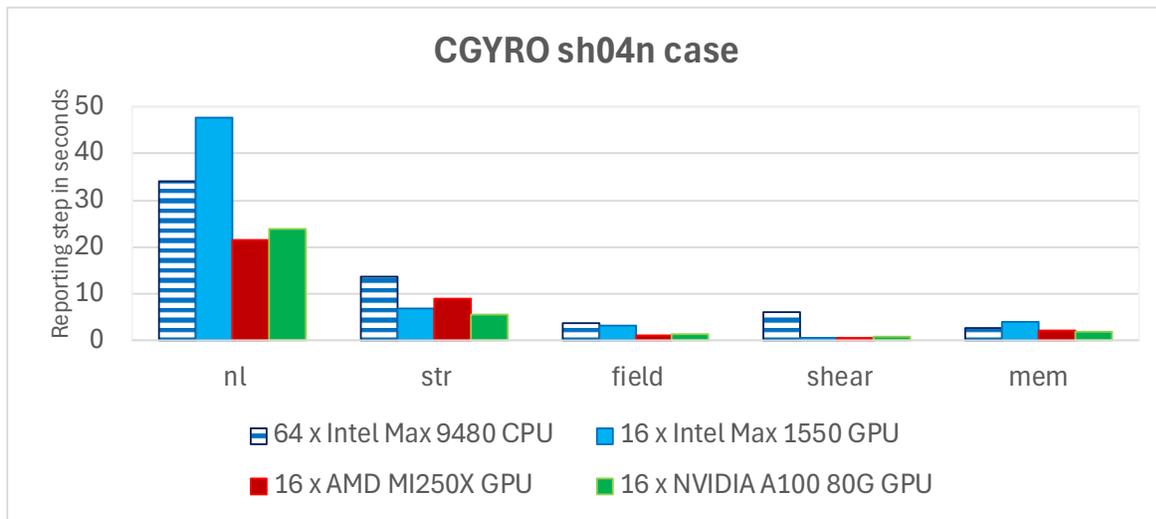

Fig B5. Comparison of measured times by code section for CGYRO `sh04n` input

Table B6. Measured time needed by reporting step for the CGYRO `bg04n` input

| #XPUs (#Nodes) | XPU type | System | nl | coll | str | field | shear | mem | io | comm |
|---|---|---|---|---|---|---|---|---|---|---|
| 64 (32) | Intel Max 9480 CPU | Stampede3 | 48.3 | 0.8 | 13.6 | 6.6 | 5.4 | 2.6 | 0.2 | 51.8 |
| 16 (4) | Intel Max 1550 GPU | Stampede3 | 42.5 | 0.4 | 6.8 | 2.9 | 0.6 | 3.0 | 0.7 | 56.3 |
| 16 (4) | AMD MI250X GPU | Frontier | 27.2 | 0.6 | 8.3 | 1.3 | 0.4 | 2.1 | 1.2 | 15.7 |
| 16 (4) | NVIDIA A100 80G GPU | Perlmutter | 23.7 | 0.2 | 5.1 | 1.4 | 0.7 | 1.9 | 1.0 | 25.6 |
| 16 (4) | NVIDIA A100 40G GPU | Perlmutter | 27.4 | 0.3 | 5.8 | 1.8 | 0.8 | 2.1 | 1.0 | 26.9 |

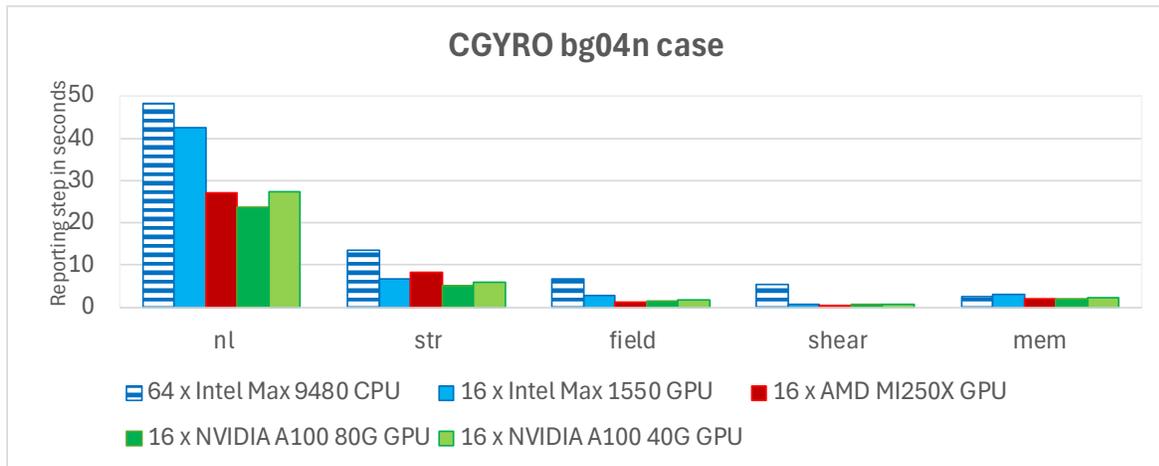

Fig B6. Comparison of measured times by code section for CGYRO`bg04n` input